\documentclass[aip, amsmath,amssymb, reprint]{revtex4-1}
\usepackage{graphicx}
\usepackage{dcolumn}
\usepackage{bm}

\usepackage[utf8]{inputenc}
\usepackage[T1]{fontenc}
\usepackage{mathptmx}
\usepackage{color}
\usepackage{siunitx}
\usepackage{soul,xcolor}

\usepackage{lineno}

\begin{document}
\setstcolor{red}
\preprint{AIP/123-QED}

\title{Making high-quality quantum microwave devices with van der Waals superconductors}

\author{Abhinandan Antony}
  \affiliation{Department of Mechanical Engineering, Columbia University, New York, NY 10027, USA}
\author{Martin V. Gustafsson}
 \affiliation{Raytheon BBN Technologies, Quantum Engineering and Computing Group, Cambridge, Massachusetts 02138, USA}
 \author{Anjaly Rajendran}
 \affiliation{Department of Electrical Engineering, Columbia University, New York, NY 10027, USA}
\author{Avishai Benyamini}
 \affiliation{Department of Mechanical Engineering, Columbia University, New York, NY 10027, USA}
\author{Guilhem Ribeill}
  \affiliation{Raytheon BBN Technologies, Quantum Engineering and Computing Group, Cambridge, Massachusetts 02138, USA}
\author{Thomas A. Ohki}
\affiliation{Raytheon BBN Technologies, Quantum Engineering and Computing Group, Cambridge, Massachusetts 02138, USA}
\author{James Hone}
 \affiliation{Department of Mechanical Engineering, Columbia University, New York, NY 10027, USA}
\author{Kin Chung Fong}%
 \email{kc.fong@raytheon.com}
\affiliation{Raytheon BBN Technologies, Quantum Engineering and Computing Group, Cambridge, Massachusetts 02138, USA}

\date{\today}

\begin{abstract}
Ultra low-loss microwave materials are crucial for enhancing quantum coherence and scalability of superconducting qubits. Van der Waals (vdW) heterostructure is an attractive platform for quantum devices due to the single-crystal structure of the constituent two-dimensional (2D) layered materials and the lack of dangling bonds at their atomically sharp interfaces. However, new fabrication and characterization techniques are required to determine whether these structures can achieve low loss in the microwave regime. Here we report the fabrication of superconducting microwave resonators using NbSe$_2$ that achieve a quality factor $Q > 10^5$. This value sets an upper bound that corresponds to a resistance of $\leq 192 \mu\Omega$ when considering the additional loss introduced by integrating NbSe$_2$ into a standard transmon circuit.  This work demonstrates the compatibility of 2D layered materials with high-quality microwave quantum devices.
\end{abstract}

\maketitle

Fault-tolerant quantum computation requires qubits with long coherence times \cite{Barends:2014fu,Chow:2014ena}. Over the past two decades, researchers have increased the lifetimes of superconducting qubits by orders of magnitude, in large part by improving their geometries and constituent materials to reduce microwave losses\cite{Devoret:2013jza,Oliver:2013dn}. In parallel with these advances, stacked van der Waals (vdW) materials have emerged as a platform of wide applicability, enabling new devices such as gate-tunable qubits \cite{Wang:2019dy} and resonators \cite{Schmidt:2018ji}, and single-photon detectors \cite{Walsh:2021il}. Because the vdW flakes are fabricated by cleaving pristine crystals, we could reduce the number of defects and contamination both on the surfaces and in the bulk, thus potentially improving the qubit coherence. However, the use of vdW superconductors such as niobium diselenide (NbSe$_2$), tungsten ditelluride (WTe$_2$), and tantalum disulphide (TaS$_2$), in quantum devices has remained largely unexplored. While the existing fabrication techniques have proven to be sufficient to achieve good DC transport measurement characteristics, the same process does not directly apply to quantum devices. For instance, the dielectric loss in microwave frequencies at the single-photon limit, rather than the DC loss measured at high power, determines the performance of superconducting qubits; a finite electrical contact resistance \cite{Telford:2018fq} is acceptable in DC transport, but detrimental to quantum devices. Moreover, superconductors in the 2D limit may have additional loss channels due to reduced superfluid stiffness \cite{benyamini2019fragility}, unpaired vortices \cite{Nsanzineza:2014cs,Khestanova:2018fsa}, as well as  spin-orbit coupling \cite{Xi:2016bu}. To use vdW superconductors in low-loss microwave circuits, it is necessary to integrate them with conventionally fabricated superconductors to produce high-quality factor resonators and long-coherence time qubits.

Here, we present the integration of exfoliated NbSe$_2$ flakes and deposited Al into a monolithic superconducting microwave platform. We verify a method to fabricate transparent contacts between the two materials and confirm that the incorporation of NbSe$_2$ in our circuit does not introduce excess microwave dissipation. This development paves the way for more complex quantum microwave circuits that take advantage of the unique features of vdW superconductors, such as compact transmon qubits\cite{Koch:2007gz,Zhao:2020jw} with stacked parallel-plate capacitors\cite{Cicak:2010fr}.

\begin{figure}[b]
\includegraphics[width=0.8\columnwidth]{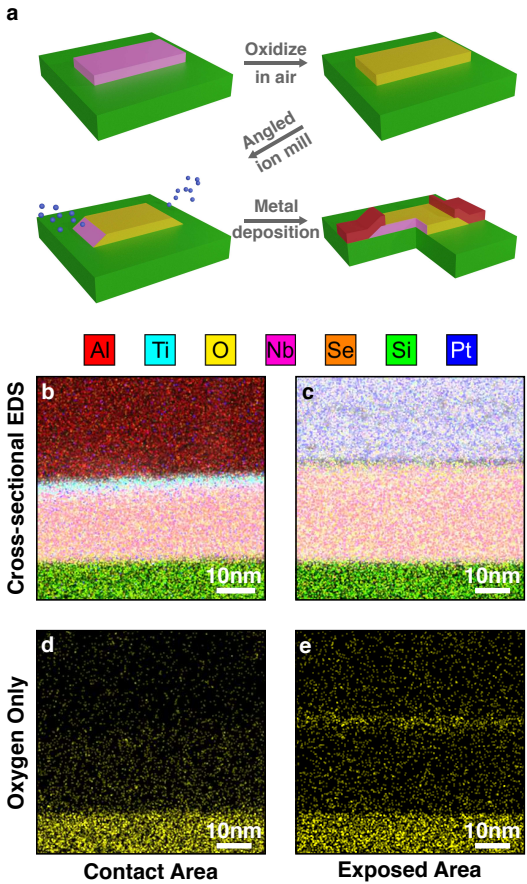}
\caption{Fabrication process and cross-sectional energy-dispersive spectroscopy (EDS). \textbf{(a)} NbSe$_2$ is transferred onto a pre-patterned void in the ground plane made of niobium on a silicon chip (top left). The surface of NbSe$_2$ is oxidized upon exposure to air (top right). Argon ion milling is performed at an angle (bottom left) exposing pristine NbSe$_2$ and contacts are then deposited in the same chamber without exposure to air (bottom right). \textbf{(b)} and \textbf{(c)} False color cross sectional EDS maps of the contacts and exposed areas, respectively. \textbf{(d)} and \textbf{(e)} Contribution of oxygen to each of the above maps.}
\label{fig:fab}\end{figure}

Our process of combining NbSe$_2$ with conventional superconductors to make high-$Q$ microwave resonators starts with a high resistivity silicon substrate sputter-coated with 200 nm of niobium. Resonators are made from the Nb film by photolithography and plasma etching. At the end of each resonator, a 400 $\mathrm{\mu}$m $\times$ 400 $\mu$m area of bare Si is left open for the placement of vdW flakes or heterostructures. The use of pre-patterned Nb films reduces the risk of getting polymer residue from the vdW materials deposition trapped between the metal film and the substrate. Inside a glove box filled with nitrogen gas, we exfoliate NbSe$_2$ flakes onto a silicon wafer with 285 nm of SiO$_2$ on the top. The thickness of NbSe$_2$ flakes are on the order of 35 nm, initially selected by their optical contrast and later confirmed with an atomic force microscope. We use a polypropylene carbonate (PPC)/Polydimethylsiloxane (PDMS) based dry-pick-up and transfer process \cite{Wang:2013ch} to transfer selected flakes onto the uncovered silicon areas on the patterned resonator substrates. During this process, care is taken to restrict the size of the contact area between the PPC/PDMS stamp and the substrate.  Finally, we use e-beam lithography to define an electrode pattern that connects the NbSe$_2$ to the Nb, and also extends the ground plane into the region near the vdW flake (Fig. 1a).

 Upon contact with air, a layer of native oxide inevitably forms on the outer surfaces of the NbSe$_2$ \cite{Cao:2015fe}. To remove this oxide layer and ensure superconducting contact, we use in-situ argon ion milling prior to metal deposition to remove the surface layers of the materials exposed through the resist mask \cite{Sinko:2021hi}. The ion energy is 500 eV and the current 30 mA, and the milling is done at an angle of 22.5$^{\circ}$ which gives a milling rate of $\sim$ 9 nm/min. We apply the ion beam for 45-60 s under continuous rotation of the sample, which makes the millilng deep enough such that both a horizontal surface and a vertical edge of the NbSe$_2$ flake are available for metal contact. Without breaking vacuum, we deposit an adhesion layer of Ti (3 nm) followed by a Al (typically 70 - 120 nm) over the freshly exposed surfaces, followed by lift-off in acetone, thereby completing the contact between Al and NbSe$_2$ as well as between Al and Nb.

 To confirm the effectiveness of this technique, we perform cross-sectional energy-dispersive spectroscopy (EDS) (Fig. 1b-e). We find that the NbSe$_2$ under the Al contact (Fig. 1b) is thinner by about 5 nm than in other areas (Fig. 1c) due to the ion milling. By highlighting the oxygen content (Fig. 1d and e), we can see a layer with a high concentration of oxygen atoms near the surface of NbSe$_2$ that has been exposed to the air, which is absent in NbSe$_2$ that has been covered by Al.
 \begin{figure}
 \includegraphics[width=0.8\columnwidth]{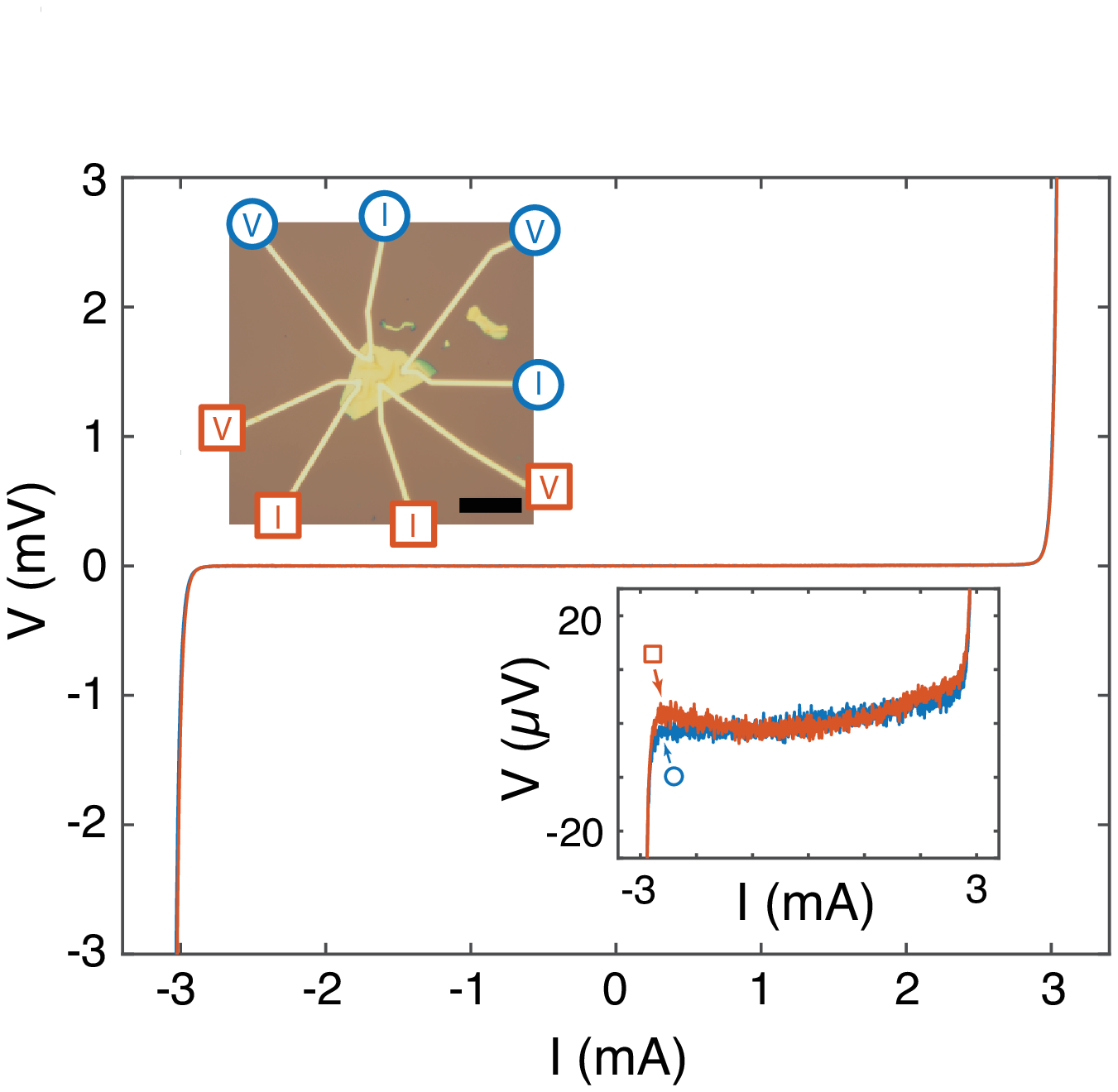}
 \caption{DC characteristics. Voltage $V$ versus bias current $I$ between two NbSe$_2$-Al contacts, showing pronounced switching to a resistive state at $I \approx 3$ mA. Data for two pairs of contacts are included in the plot (see upper inset) and they fall on top of each other. The lower inset shows a zoom-in (blue and orange), with a constant voltage offset subtracted from each. The symbols correspond to the contact configurations in the upper inset. For currents below the transition threshold, we observe no signs of electrical resistance. The drift near zero voltage can be attributed to limitations in the DC instrumentation. Top-left inset: optical image of the device used in the DC measurement. The black scale bar is 20 $\mathrm{\mu}$m long.}
 \label{fig:DC}\end{figure}

 To confirm that there is an electrical contact to NbSe$_2$, we fabricate test devices for DC measurement using the fabrication procedures described above (Fig. 2 top-left inset). Measuring the resistance through two contacts in series at a temperature of 25 mK, we find no observable resistance in the I-V characteristic for currents lower than $\sim$3 mA. Below this threshold, we measure a variation in voltage lower than 10 $\mathrm{\mu V}$, and attribute the drift to limitations in our instrument (bottom-right inset). The sharp voltage rise at $\sim$3 mA can be due to the bias current exceeding the critical current density of Al superconductor electrodes. For a \SI{107}{\nm} thick, \SI{2}{\um} wide Al electrode, the transition would correspond to a critical current density of \SI{14}{\giga\ampere/\square\meter}, smaller than but consistent with the value from other reports \cite{Romijn:1982bu}. Because of this, we cannot attribute the switching behavior to the Al-NbSe$_2$ contacts. 

 \begin{figure}
 \includegraphics{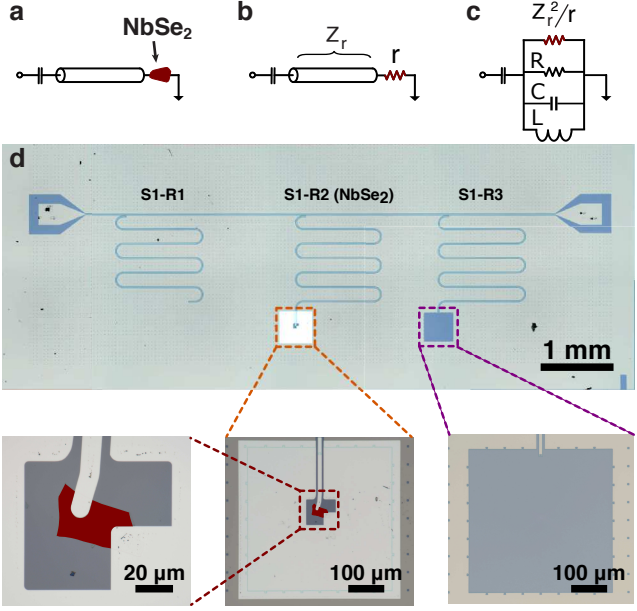}
 \caption{ Circuit diagram and optical micrographs. \textbf{(a)} Schematic of the circuit of the quarter-wave resonator terminated to ground with a NbSe$_2$ flake. \textbf{(b)} Model representation of the loss caused by the NbSe$_2$ flake and its contacts. \textbf{(c)} Equivalent circuit for the terminated resonator shown in (b) \textbf{(d)} Optical micrographs of sample S1, showing the resonators R1 (conventional, left), R2 (quarter-wave with NbSe$_2$, middle) and R3 (conventional, right), and enlarged images of the R2 NbSe$_2$ termination and R3. The NbSe$_2$ flake is highlighted in red. The termination of resonator R2 on sample S2 is shown in Fig. 6 (Appendix).}
 \label{fig:res}\end{figure}

By testing superconducting resonators with embedded NbSe$_2$, we are able to verify that the NbSe$_2$/Al system is suitable for quantum microwave devices. Since the non-NbSe$_2$ parts of the resonators are fabricated in a process that is capable of yielding high quality factors, the superconducting resonators can serve as sensitive probes for any microwave losses introduced by the NbSe$_2$ flake or its contacts\cite{Quintana:2014jp, McRae:2020hha}, which manifest as a degradation of the resonator's internal quality factor $Q_i$. Fig. 3a shows a schematic of the resonator, with the NbSe$_2$ shunting one end of a capacitively coupled coplanar waveguide (CPW) to ground, thus forming a quarter-wave resonator. The position of the NbSe$_2$ flake at the end of the resonator puts it at a maximum of the current in the resonator. We can model the loss associated with the NbSe$_2$ flake as a resistor $r$ (Fig. 3b) and represent the entire quarter-wave resonator as a parallel $RLC$ circuit with $R$, $L$, and $C$ representing the resistive loss, inductance, and capacitance of the CPW resonator (Fig. 3c). For a characteristic CPW impedance $Z_r$ and resonance frequency of $\omega_0/2\pi$, $Q_i$ is given by:
\begin{eqnarray}
Q_i &=& \omega_0\left(\frac{1}{R}+\frac{r}{Z_r^2}\right)^{-1}C
\label{eq:newQ1}\\
\frac{1}{Q_i} &=& \frac{1}{Q_0}+\frac{r}{\omega_0CZ_r^2}\label{eq:newQ2}
\end{eqnarray}
Here, $Q_0$ is the internal quality factor of the CPW part of the resonator, modeled by the resistance $R$. In this circuit model, the loss of the NbSe$_2$ flake acts as a shunt resistance of value $Z_r^2/r$.

 Fig. 3d shows optical images of one of our devices, with three meandered CPW resonators capacitively coupled to a CPW feed line running through the center of the chip. The leftmost resonator (R1) is made entirely from Nb and serves as a benchmark for the fabrication process and the experimental setup. For the two resonators to the right, the Nb patterns are geometrically identical and end in open areas of bare Si where a vdW flake can be placed. In one of these two resonators (R2, NbSe$_2$), we short the end of CPW to ground through a flake of NbSe$_2$ to create a quarter-wave resonator with center frequency $\omega_{\lambda/4}$. The other one (R3) is left unmodified and serves as a control resonator with resonance frequency $\omega_{\lambda/2} \approx 2 \omega_{\lambda/4}$.

 \begin{figure}
 \includegraphics{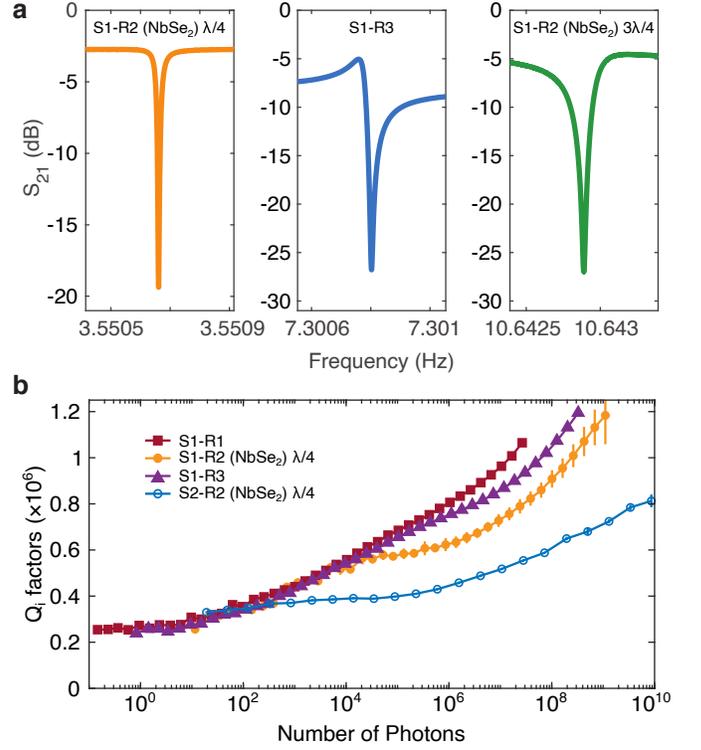}
 \caption{High frequency loss measurements \textbf{(a)} Transmitted amplitude vs frequency for on sample S1 for fundamental (left, orange) and first harmonic (right, green) modes of the quarter wave resonator (R2, NbSe$_2$), and the control resonator R3 (middle, blue). \textbf{(b)} Internal quality factor vs the number of photons in the resonator for the NbSe$2$ resonators (R2) for two samples (S1 and S2), as well as for the two conventional resonators from one sample for comparison (S1-R1 and S1-R3). Error bars corresponding to the uncertainty in the circle fitting are shown where these are larger than the data markers.}
 \end{figure}

Fig. 4a shows the resonances of the quarter-wave (R2, NbSe$_2$) and control (R3) resonators of sample S1 as notches in the transmission spectrum through the CPW feedline. The fundamental mode of the quarter-wave and control resonators are at $\omega_{\lambda/4}/2\pi$ = 3.551 GHz and $\omega_{\lambda/2}/2\pi$ = 7.301 GHz, respectively. The resonance frequency of the control resonator is slightly higher than twice the quarter-wave resonance frequency because of the Al extension that couples the NbSe$_2$ and the Nb CPW. The resonance near $3\omega_{\lambda/4}/2\pi$ = 10.643 GHz is the third harmonic of the quarter-wave resonator R2. Using the circle-fitting method \cite{Khalil:2012jr,Probst:2015dt}, as implemented in the software toolkit presented in \citet{Probst:2015dt}, we extract $Q_i$, the coupling quality factor $Q_c$, and from these values we get the applied power $P_{1\mathrm{ph}}$ that corresponds to a single photon in the resonator. We do this for each resonance as a function of applied power, and use $P_{1\mathrm{ph}}$ along with a room temperature calibration of the microwave line attenuation to determine the number of photons in the resonator for a given power \cite{Burnett2018}. The total line attenuation ranges between 70 dB and 84 dB depending on resonance frequency, with 30 dB mounted at the sample stage of the refrigerator (see Fig. 5 in the Appendix). Fig. 4b shows $Q_i$ for each NbSe$_2$ resonator from two different samples (S1-R2 and S2-R2) as well as the two conventional resonators on of sample (S1-R1 and S1-R3) for comparison. We find that all the tested resonators have similar loss in the single-photon limit and follow similar increasing trends of $Q_i$ with photon number. This is indicative of the low-power loss being dominated by two-level systems, though we cannot rule out contributions from other mechanisms, such as quasiparticle dissipation\cite{Patel:2017fz,Grunhaupt:2018ku,McRae:2020hha, Goppl:2008iu}. At high power, $Q_i$ differs between samples, which can be caused by coupling to different TLS ensembles and dissipative box modes due to differences in the resonator geometry near the NbSe$_2$ flake (see Fig. 6 in the appendix). The effectiveness of the radiation shielding can also differ somewhat between experiments, which affects the contribution to the loss by quasiparticles.

An all-superconducting materials platform that incorporates vdW materials may have important applications for quantum information processing. A particularly appealing prospect is to take advantage of the crystalline NbSe$_2$ to fabricate superconducting qubits \cite{antony2021q, wang2021hbn}, in which the capacitor consists of vdW superconductors with an insulating vdW dielectric sandwiched between them, and where the Josephson junction that shunts the capacitor is fabricated from Al with conventional lithographic techniques. With the dielectric made of hexagonal-boron nitride (hBN), flakes with thicknesses of a few tens of nanometers and lateral sizes of a few micrometers can be used to create parallel-plate capacitances with values in the range of typical transmons ($\sim 60$ fF). The assembly of such vdW heterostructures is well within the capability of current stacking techniques, and a qubit with this design can be orders of magnitude more compact than conventional planar transmons. The ultra-clean and crystalline interfaces of the stacked vdW flakes can potentially exhibit very low dielectric loss, which otherwise typically limits the performance of transmon qubits.

To put an upper bound on any loss caused by the NbSe$_2$, we use the worst case scenario by setting $1/Q_0$ to zero when applying Eqn. \ref{eq:newQ2}. In the low-power regime, $Q_i$ = 260 000, corresponding to an upper bound of $r = 192$ $\mathrm{\mu\Omega}$. To estimate the effect of such a dissipative element in a superconducting qubit, we consider a transmon as described above, modified with an added resistance $r$ in series with the capacitance, such that the $T_1$ relaxation time can be estimated by $T_1 = (\omega_q^2\,r\, C_q)^{-1}$ with $\omega_q/2\pi$ and $C_q$ being the frequency and capacitance of the qubit. Using a typical transmon capacitance, $C_q = $60 fF and $\omega_q = 2\pi\times$5 GHz, respectively, we estimate that $T_1$ can reach nearly 100 $\mu$s when limited by $r$. This conservative estimate confirms that the NbSe$_2$/Al system can be used to make superconducting quantum devices with long coherence times. We expect that the same fabrication and validation approach can produce high-quality quantum microwave device based on other superconducting vdW materials.

\textbf{Acknowledgements.} We thank B.~Hunt and L.~Ranzani for useful discussions. This work was supported by Army Research Office under Cooperative Agreement Number W911NF-18-C-0044. A.~A. thanks the supplemental support from QISE-NET under NSF DMR-1747426.

\subsection{Appendix}
\subsubsection{\label{sec:para}Parameters of the quarter-wave resonators}
\begin{table}[h]
\centering
\begin{tabular} {l c c c c}
\hline
Parameters & S1-R1 & S1-R2 & S1-R3 & S2-R2  \\
& & (NbSe$_2$) & & (NbSe$_2$) \\
\hline
Resonator type& $\lambda/2$ & $\lambda/4$ & $\lambda/2$ & $\lambda/4$ \\
Center frequency, $\omega_0/2\pi$ (GHz)& 8.482 & 3.550 & 7.301 & 3.280 \\
Length of transmission line, $\ell$ ($\mu$m)& 7009 & 8310 & 8153 & 8153  \\
Characteristic impedance & 50 $\Omega$ & 50 $\Omega$ & 50 $\Omega$ & 50 $\Omega$  \\
Equivalent circuit inductance (pH)& 600 & 290 & 690 & 310 \\
Equivalent circuit capacitance (fF)& 590& 700 & 680 & 760 \\
Coupling quality factor $Q_c$ & 23000 & 212000 & 200000 & 247000 \\
\hline
\end{tabular}
\caption{\textbf{List of resonator parameters.}}
\label{tab:para}
\end{table}

\subsubsection{Additional figures}
 \begin{figure}[h]
 \includegraphics[keepaspectratio, width=0.6\columnwidth]{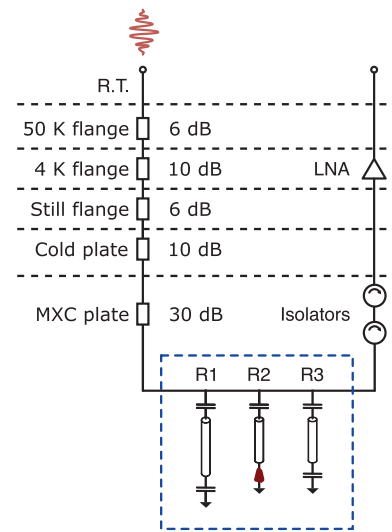}
 \caption{Experimental setup. The total attenuation exceeds that of the fixed attenuators shown at different temperature stages due to the contribution of dissipative cables. The frequency dependent total attenuation, measured at room temperature, is used to determine the resonator photon number.}
 \end{figure}

 \begin{figure}[h]
 \includegraphics[keepaspectratio, width=0.5\columnwidth]{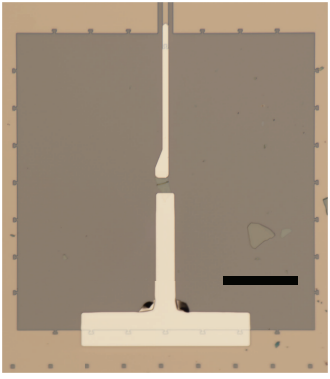}
 \caption{Optical image of the NbSe$_2$ termination of the S2-R2 resonator. The scale bar is 100 $\mu$m long.}
 \end{figure}
 
 \begin{figure}
 \includegraphics[width=\columnwidth]{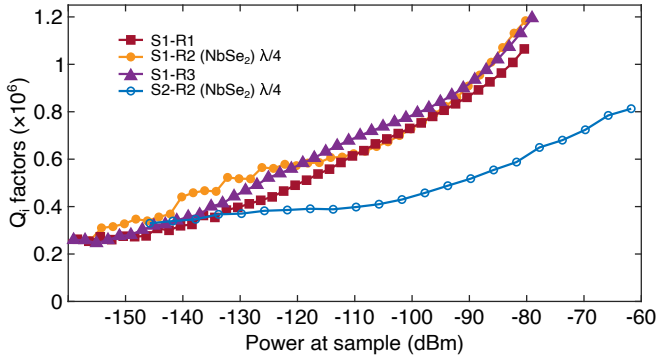}
 \caption{Internal quality factor vs power applied to the sample for the fundamental mode of NbSe$_2$ resonators S1-R2 and S2-R2, as well as conventional resonators S1-R1 and S1-R3 for comparison.}
 \end{figure}

 \begin{figure}
 \includegraphics[width=\columnwidth]{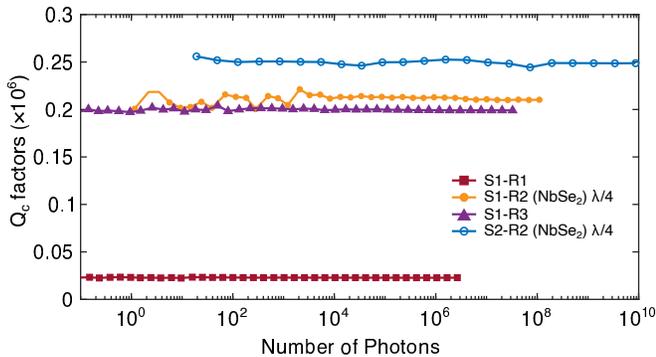}
 \caption{Coupling quality factor vs the number of resonator photons for the fundamental mode of NbSe$_2$ resonators S1-R2 and S2-R2, as well as conventional resonators S1-R1 and S1-R3 for comparison.}
 \end{figure}

\bibliography{QuarterV6}

\end{document}